\documentclass{iaus}
\usepackage{graphicx}

\title[IAU266.~~Intermediate-mass black holes in star clusters] 
{{\sl HST}'s Hunt for Intermediate Mass Black Holes in Star Clusters}

\author[J. Chanam\'e et al.]   
{Julio Chanam\'e$^1$,
 Justice Bruursema$^2$, Rupali Chandar$^3$, Jay Anderson$^4$, Roeland
 van der Marel$^4$, \& Holland Ford$^2$}

\affiliation{$^1$Department of
  Terrestrial Magnetism, Carnegie Institution of Washington,
  Washington, DC, USA \\ email: {\tt jchaname@dtm.ciw.edu} \\[\affilskip]
$^2$Department of Physics and Astronomy, Johns Hopkins University,
Baltimore, MD, USA
\\[\affilskip]
$^3$Department of Physics and Astronomy, The University of Toledo,
Toledo, OH, USA \\[\affilskip]
$^4$Space Telescope Science Institute, Baltimore, MD, USA \\[\affilskip]
}

\pubyear{2010}
\volume{266}  
\jname{Star clusters: basic galactic building blocks}
\editors{R. de Grijs \& J. R. D. L\'epine, eds.}
\begin{document}

\def\kms{{\rm \,km}\,{\rm s}^{-1}}
\def\masyr{{\rm \,mas}\,{\rm yr}^{-1}}

\maketitle

\begin{abstract}
  Establishing or ruling out, either through solid mass measurements
  or upper limits, the presence of intermediate-mass black holes
  (IMBHs; with masses of $10^2 - 10^5$ M$_{\odot}$) at the centers of
  star clusters would profoundly impact our understanding of problems
  ranging from the formation and long-term dynamical evolution of
  stellar systems, to the nature of the seeds and the growth
  mechanisms of supermassive black holes.  While there are sound
  theoretical arguments both for and against their presence in today's
  clusters, observational studies have so far not yielded truly
  conclusive IMBH detections nor upper limits.  We argue that the most
  promising approach to solving this issue is provided by the
  combination of measurements of the proper motions of stars at the
  centers of Galactic globular clusters and dynamical models able to
  take full advantage of this type of data set.  We present a program
  based on {\sl HST} observations and recently developed tools for
  dynamical analysis designed to do just that.

\end{abstract}

\firstsection 
\section{Introduction}

There is solid observational evidence for the existence of black holes
(BHs) in two very different regimes.  Stellar-mass BHs, with masses in
the range $5-15$ M$_\odot$, are the end products of normal high-mass
stellar evolution. Solid mass estimates for these BHs come from
measurements of the mass function in several tens of Galactic X-ray
binaries. Supermassive BHs (SMBHs), at the other extreme, with masses
between $10^{5.5}$ and $10^{9.5}$ M$_\odot$, have been reliably
weighed in galactic centers based on the motions of their surrounding
stars and gas. A tight correlation between the masses of SMBHs and the
velocity dispersion of their host spheroid (the $M_{\rm BH}-\sigma$
relation) evidences that an intimate link must exist between these BHs
and the processes relevant to galaxy formation, and highlights the
importance of understanding the seeds and growth mechanisms of
SMBHs. IMBHs, with masses in between those of the above two regimes,
are likely to play a key role in these questions but they are yet to
be convincingly detected.

Existing theoretical work provides arguments for and against the
presence of IMBHs at the centers of globular clusters (GCs).  Several
possible channels of formation have been proposed (see van der Marel
2004 for a review), and, interestingly, all those mechanisms lead to
IMBH masses consistent with an extrapolation of the $M_{\rm
BH}-\sigma$ relation down to dispersions typical of GCs.  On the other
hand, it has been argued that it may be difficult for GCs, given their
relatively shallow potential wells, to retain growing massive black
holes at their centers for too long (Baker et al. 2008;
Holley--Bockelmann et al. 2008; Moody \& Sigurdsson 2009).

This current debate, however, should not prevent astronomers to set
out plans to look for the existence of these objects.  To the question
of whether IMBHs are possible or not in GCs, the right attitude should
be one that, although we do not know the answer at this moment, we
will also act independently from theoretical prejudices and
observationally probe for their presence anyway.  Any tight
upper limits to the masses of central compact objects in a number of
GCs are equally important for the field as any solid mass measurements
eventually confirming their existence.  Certainly, any of these
alternatives would clearly constitute an important improvement over
today's situation.

\section{Current state of the hunt for IMBHs in GCs}

As in the case of SMBHs in active galactic nuclei, one could argue
that the most unambiguous signature of the presence of a massive
compact object would be the detection of central, unresolved radio
and/or X-ray luminosity generated by the accretion of matter onto the
central object.  In the case of star clusters, however, one must make
sure that this signature is not consistent with an accumulation of
more common stellar-mass accreting objects (cataclysmic variables,
X-ray binaries, stellar-mass black holes) or pulsars.  This danger is
well illustrated by the case of a GC associated with the giant
elliptical galaxy NGC 4472 in the Virgo cluster, where Maccarone et
al. (2007) reported a strong and highly variable X-ray signature
``which rules out any object other than a black hole in such an old
stellar population'', then estimating about 400 M$_\odot$ for the mass
of this object.  However, as noted by those authors, such mass
estimate was necessarily based on several assumptions regarding highly
uncertain variables of accretion physics (e.g., innermost stable
orbit, accretion rate, disk and continuum model, unabsorbed
luminosity, etc.), and can hardly be considered a mass measurement.
Indeed, they later determined, based on optical spectroscopy, that the
object in question is most likely a stellar-mass black hole (Zepf et
al. 2008).

As a consequence, while the observation of central, unresolved
accretion signatures originating from star clusters certainly
constitutes evidence for the possible presence of IMBHs, and,
moreover, may be considered a prime way to search for IMBH candidates,
it is nevertheless not adequate for the purposes of a solid,
unambiguous mass measurement.  The same considerations apply to the
possibility of catching an event of stellar disruption by an IMBH
(Irwin et al. 2009).

Therefore, for unambiguous mass measurements, we need to rely on
dynamics, and there have been various observational efforts in this
direction. Based on dynamical modeling of measured line-of-sight (LOS)
kinematics, evidence for the presence of IMBHs has been claimed in
three GCs: M15 and $\omega$ Cen in the Milky Way (Gerssen et al. 2002
and Noyola et al. 2008, respectively), and G1 in M31 (Gebhardt et
al. 2002, 2005). In the cases of M15 and G1, however, the available
data are also consistent with nonequilibrium models in which mass
segregation occurs over time, but which do not posses an IMBH.
(Baumgardt et al. 2003a,b).  Moreover, those measurements are
restricted to the luminous red giants in the clusters.  Not being very
numerous, any analysis based on the kinematics of the giants will be
more affected by shot noise and projection effects (i.e., not knowing
whether the tracers contributing to most of the velocity signal are
actually probing the IMBH's sphere of influence, or whether they are
instead located in front of or behind the cluster center) than if the
data set were dominated by the more numerous, although fainter,
main-sequence stars.

The cases of G1 and $\omega$ Cen are unique in their own way.  We
start by pointing out that, since the subject of this meeting is star
clusters, and although the presence of an IMBH does not have anything
to do with this fact, it is nevertheless appropriate to recall that
$\omega$ Cen and G1 are both likely to be the nuclei of stripped dwarf
galaxies rather than regular GCs.  However, as this is yet to be
considered an established result, we will discuss them here anyway
and, instead of having to specify their uncertain nature by calling
them `stellar spheroids of low central velocity dispersion', we will
refer to them loosely as clusters.  With respect to G1, one can argue
that this remains the best case so far for an IMBH in a GC.  This is
because, on top of the dynamical evidence quoted above, this cluster
is also a source of radio and X-ray emission (Ulvestad et al. 2007;
Kong 2007), as one would expect for an accreting compact object.  The
problem with this additional accretion evidence, however, is one of
angular resolution, since both the radio and X-ray source positions
are too uncertain and they could still be the result of a number of
less exotic objects, such as X-ray binaries or CVs distributed
throughout the cluster.  Of course, this situation is expected to
improve when performing higher-angular-resolution
observations.\footnote{Kong et al. (2009) recently reported
high-resolution observations with {\sl Chandra}, locating the X-ray
emission within the core radius of G1. Based on the ratio
of X-ray to Eddington luminosity, the authors suggest the emission is
more likely coming from low-mass X-ray binaries.} In the case of
$\omega$ Cen, the problem has been again related to the effect of
bright giants, in this case the effect that their shot noise has on
the determination of the cluster center, a key step in the
determination of light and velocity-dispersion profiles.  Briefly,
Anderson \& van der Marel (2009) used high-angular-resolution {\sl
HST} images to measure the proper motions of cluster members, most
still on the main sequence, and found a cluster center about 12 arcsec
away from that determined by Noyola et al.  (2008).  The light and
(2D) velocity-dispersion profiles computed with respect to the new
center did not show the signatures that Noyola et al.  (2008)
atributed to a 40\,000 M$_\odot$ IMBH.  Next, van der Marel \&
Anderson (2009) constructed dynamical models with the proper-motion
data and determined that models both with and without a black hole fit
the data equally well, reporting a tight upper limit of 12\,000
M$_\odot$ in case an IMBH is present.

\section{Current Theoretical Understanding}

Until just a few years ago, the generalized intuition indicated that
the best places to look for IMBHs were the centers of very dense
stellar systems such as the cores of Galactic globular clusters with
very steep luminosity profiles. Solid theoretical basis for this
expectation came from the seminal work of Bahcall \& Wolf (1976,
1977), which predicted the development of stellar density cusps
surrounding massive central BHs. In well-relaxed clusters, groups of
stars of different mass formed cusps with correspondingly different
slopes, reflecting the well-known phenomenon of mass segregation. This
prompted the selection of Galactic globular clusters with dense,
collapsed cores for observational studies that would look for the
dynamical signature of an IMBH at their centers. However, drawing
conclusions from the simulations as to what to expect observationally
has never been an easy task, given that while the theoretical results
are unambiguous regarding the mass of the particles, the observations
can only track the luminous stars, and conversion from one to the
other depends on many assumptions. More recently, it was realized that
an IMBH at the center of a star cluster constitutes an additional
source of heat acting against the rapid collapse of the cluster's
core. Baumgardt et al. (2004a,b) found that the IMBH not only prevents
the collapse of the core but also produces an expansion of the
cluster, making it unlikely that dense core-collapsed clusters could
harbor IMBHs. Therefore, a completely new picture of where it would be
best to look for IMBHs has emerged: unlike collapsed cores and dense
central regions, the most likely clusters to harbor an IMBH would be
those whose projected surface brightness profile are well fit by
regular King-like models with an extended core and intermediate
concentrations ($W_0 \sim 7, c \sim 1.5$), with the surface brightness
rising slightly only in the very inner regions of this core, forming a
shallow, barely perceptible density cusp in the form of a power law
with slope $\alpha = {\rm d}\log {\rm SB}/{\rm d}\log r \sim -0.25$
(Baumgardt et al. 2005; Miocchi 2007; Trenti et al. 2007).

\section{Towards Solid Mass Measurements or Upper Limits}

In Section\,2 we started by arguing that the observation of accretion
signatures, while great for the detection of IMBH candidates, was not
adequate for a reliable mass measurement of satisfactory precision,
for which we had to rely on dynamical analyses.  The current status of
the evidence for IMBHs in GCs, however, proves that dynamics has its
own problems.  A large part of these, as summarized in Section\,2,
have to do with the limited amount of data available for those
studies, which did not provide enough kinematic information to, say,
distinguish the effect of a single massive central object from that of
a more extended mass distribution.  But this constitutes just one
aspect of the potential problems associated with dynamics, the other
side having to do with the subsequent handling of the available data.
Indeed, in the process of the modeling of kinematic data, two of the
most important dangers are (i) the inadequate exploitation of the full
information content of a given dataset and (ii) the adoption of
simplifying assumptions that are too restrictive and thus explore only
a limited region of the space of possible solutions.

The management of these latter points can have profound implications
for the inferred mass distribution (arguably the main goal of stellar
dynamics), thus creating a delicate balance between the available
observations for a given problem and the complexity of the models
chosen to make use of such data.  This balance was nicely exemplified
by the initial debate regarding the dark-matter content of some
intermediate-luminosity elliptical galaxies, during which LOS velocity
measurements of planetary nebulae in the halos of these
galaxies---showing a Keplerian-like decline of velocity dispersion
with radius---were interpreted as evidence of the presence of little,
if any, dark matter in those galactic halos (Romanowsky et al. 2003).
Although this remains a subject of current discussion, it was later
shown that such conclusions may well be the result of too restrictive
assumptions regarding both the geometry of the underlying potential
and the anisotropy of the orbits of the planetary nebulae around those
galaxies (Dekel et al. 2005; de Lorenzi et al. 2009).

\subsection{The Need for HST}

The key for an unambiguous assessment of the presence of IMBHs in GCs
is availability of many stars with well-measured velocities probing
the GC's core, and containing enough kinematic information to permit
models to be very general (Section\,4.2). This demands:

\begin{itemize}

\item {\it High spatial resolution:} stellar velocities need to be
  measured inside the sphere of influence of the putative BH,
  $r_{\rm BH} \simeq 0.39'' (M_{\rm BH}/10^3\,{\rm
  M}_{\odot})(10\kms/\sigma)^2(10\,{\rm kpc}/D)$. Thus, for an IMBH of
  $3\times10^3$ M$_\odot$, many stars have to be well resolved inside
  $r_{\rm BH} \sim 1''$.

\item {\it At least two velocity components:} a well-known degeneracy
  between mass and aniso\-tropy (Binney \& Mamon 1982) is the main
  obstacle for a reliable measure of the mass distribution in stellar
  systems based on line-of-sight velocities.  Proper motions provide
  two components of the space velocity, allowing to actually measure
  the anisotropy of the stellar orbits instead of assuming it.

\item {\it Optimization of observing (wavelength) window:} high levels
  of astrometric accuracy require minimizing as much as possible the
  red light from nearby, luminous giants and the unresolved background
  of low-mass stars populating the crowded cores of GCs. Thus,
  observations must be conducted at short enough wavelengths, from the
  blue to the near-ultraviolet.

\end{itemize}

Even with the largest-aperture telescopes and assisted by adaptive
optics, ground-based observations are unable to produce radial
velocities and/or proper motions for many stars in the inner arcsecond
of GCs (e.g., Gebhardt et al. 2000). Built exactly to overcome these
problems, {\sl HST} is by design the {\it only} instrument that, by
largely satisfying the three above requirements, will be able to
resolve, within the next decade, the question of the existence of
IMBHs in GCs.

\subsection{The Need for Specialized Modeling}

Appropriate data represent only half of what is needed, and highly
specialized models are required to avoid the shortcomings of more
simplified methods that were developed for other circumstances.
First, special machinery is necessary to handle data of a discrete
nature (i.e., individual 2D velocities of a number of resolved
stars). The most typical approaches for dynamical analysis (the Jeans
equations and orbit-based models designed to work with observed
velocity profiles from integrated light) must necessarily bin the data
to produce averaged profiles of velocity dispersions (see figure 13 in
van de Ven et al. 2006).  Clearly, binning amounts to not exploiting
the entire information content of the dataset, and, since only a
limited number of stars are available inside the BH's sphere of
influence, this can only have the effect of degrading and possibly
erasing the signature of an IMBH.

Second, given that the signature of any central BH would be imprinted
on the detailed orbital structure of the nearest stars, overly
restrictive assumptions regarding the form of the stellar distribution
function, most crucially the degree of isotropy of the orbits of the
tracer stars, must be avoided.  This point becomes obvious when
considering that, for M15 for example, {\sl HST} LOS velocities
indicate that there is a clear increase in the net rotation of the
stars in the inner arcseconds of the cluster (see figure 9 of Gerssen
et al. 2002).  This fact stands in stark contradiction to the standard
expectation that relaxation should rapidly drive the velocity
distribution at the centers of GCs towards isotropy, thus illustrating
the necessity of using the most general models possible to fit these
kinematics.

Third, the fact that there is flattening and substantial rotation seen
in some GCs makes the use of axisymmetric models, rather than
spherical ones, a must.

\section{Our ongoing HST program}

Starting around 2004, our team embarked on a comprehensive program to
use accurate {\sl HST} proper-motion measurements of the stars in the
central regions of a number of Galactic GCs and search for evidence of
IMBHs.  Reflecting the theoretical expectations of that time, all GCs
included in these initial {\sl HST} programs were selected to be among
those with high central densities (several of them being
core-collapsed systems).  Through the following years, target clusters
spanning a variety of structural properties were included in the
sample, and today we have accumulated data, from our own programs as
well as by exploiting the {\sl HST} Archive, for a total of nine GCs
with two or more astrometric epochs, producing time baselines between
2 and 6 years.  Reflecting the history of {\sl HST} of the last few
years, the data were obtained using those instruments providing the
highest angular resolution possible at the time of the observations.
Table 1 summarizes our current dataset.  We are currently in the
process of reducing these data, and preliminary stellar proper-motion
catalogs for some of these clusters are being generated (Bruursema et
al., in prep.).  For a review of this intensive process, see the
proper-motion catalog of $\omega$ Cen of Anderson \& van der Marel
(2009).

We plan to perform the dynamical analysis of all these data sets by
making use of Schwarzschild models specifically designed to exploit
data of a discrete nature, such as that of our proper-motion catalogs.
Schwarzschild's technique is arguably the most developed and
well-tested method available for constraining the detailed mass
distribution of equilibrium stellar systems, and is based on the
simple idea of finding the best combination (or superposition) of all
possible orbits that, allowed by some previously specified potential,
reproduces both the spatial distribution of the tracers (i.e., the
light distribution) and the measured kinematics.  The simplicity and
success of the method therefore rely on two aspects: (i) that the
overall stellar system can be considered to be in equilibrium (a safe
assumption in the case of old Galactic GCs) and (ii) whether or not
the set of orbits considered for the superposition is really
comprehensive.  As long as these two conditions are satisfied, the
method is very general and free from most assumptions.  As of today,
the only orbit-based dynamical tool available that fulfills all
requirements outlined in Section\,4.2 is the discrete Schwarzschild
code we recently developed and tested (Chanam\'e et al. 2008), and
which we will be employing in this program.

\begin{acknowledgments}

  JC acknowledges the support of the American Astronomical Society and
  the International Astronomical Union in the form of travel grants to
  attend the IAU General Assembly, as well as the support from NASA
  through Hubble Fellowship grant HF-01216.01-A, awarded by the Space
  Telescope Science Institute, which is operated by the Association of
  Universities for Research in Astronomy, Inc., under NASA contract
  NAS5-26555.

\end{acknowledgments}
 
\begin{table}
  \begin{center}
  \caption{Details of our {\sl HST} program.}
  \label{tab1}
 {\scriptsize
  \begin{tabular}{|l|c|c|}\hline 
{Target GC} & {Instruments} & {Program ID} \\
& {Time baseline} & {PI} \\ \hline
NGC 2808 & & GTO-10335 \\
NGC 6341 (M92) & HRC/HRC & GTO-11801 \\
NGC 6752 & 2 years$^1$ & Ford \\ \hline
NGC 362 & & GO-10401 \\
NGC 6624 & & GO-10841 \\ 
NGC 6681 (M70) & HRC/WFPC2& GO-11988 \\
NGC 7078 (M15) & 4.5 years& Chandar \\
NGC 7099 (M30) & & \\ \hline
NGC 6266 (M62) & WFC/WFC3 & GO-11609 \\
& 6 years & Chanam\'e \\ \hline 
  \end{tabular}
  }
 \end{center}
\vspace{1mm}
 \scriptsize{
  $^1$These clusters will have a third epoch with WFC3 in 2010, thus
  increasing the baseline to 5--6 years.} \\
\end{table}

\end{document}